\documentclass[useAMS,usenatbib]{mn2e}
\usepackage{graphicx}
\usepackage{epsfig}
\usepackage{subfigure}

\def\farcs{\hbox{$.\!\!^{\prime\prime}$}}

\def\mnras{MNRAS}

\def\aap{A\&A}

\def\pasp{PASP}

\title[Wavelength Self-Calibration and Sky Subtraction for Fabry-P\'erot Interferometers]{Wavelength Self-Calibration and Sky Subtraction for Fabry-P\'erot Interferometers: Applications to OSIRIS}

\author[Weinzirl~et.~al.]{Tim Weinzirl$^{1}$\thanks{E-mail:
timothy.weinzirl@nottingham.ac.uk}, Alfonso~Arag\'on-Salamanca$^{1}$, Steven~P.~Bamford$^{1}$,
\newauthor Bruno~Rodr\'iguez~del~Pino$^{1,2}$, Meghan~E.~Gray$^{1}$, Ana~L.~Chies-Santos$^{3,4}$\\
$^{1}$School of Physics and Astronomy, The University of Nottingham, University Park, Nottingham, NG7 2RD, UK\\
$^{2}$Centro de Astrobiolog\'ia, INTA-CSIC, Madrid, Spain\\
$^{3}$Departamento de Astronomia, Instituto de Física, Universidade Federal do Rio Grande do Sul, Porto Alegre, R.S, Brazil\\
$^{4}$Departamento de Astronomia, Instituto de Astronomia, Geof\'isica e Ci\^encias Atmosf\'ericas, Universidade de Sa\~o Paulo, S\~ao Paulo, SP, Brazil\\
}
\begin{document}

\date{Accepted . Received ; in original form }

\pagerange{\pageref{firstpage}--\pageref{lastpage}} \pubyear{2015}

\maketitle

\label{firstpage}

\begin{abstract}
We describe techniques concerning wavelength calibration and sky subtraction to maximise the 
scientific utility of data from tunable filter instruments. While we specifically address 
data from the Optical System for Imaging and low Resolution Integrated Spectroscopy instrument 
(OSIRIS) on the 10.4~m Gran Telescopio Canarias telescope, our discussion is generalisable to 
data from other tunable filter instruments. A key aspect of our methodology is a coordinate 
transformation to polar coordinates, which simplifies matters when the tunable filter data is 
circularly symmetric around the optical centre. First, we present a method for rectifying 
inaccuracies in the wavelength calibration using OH sky emission rings. Using this technique, 
we improve the absolute wavelength calibration from 
an accuracy of $\sim5$~$\rm\AA$ to 1~$\rm\AA$, equivalent to $\sim7\%$ of our instrumental 
resolution, for 95\% of our data.   Then, we discuss a new way to estimate the background sky 
emission by median filtering in polar coordinates. This method 
suppresses contributions to the sky background from the outer envelopes of distant galaxies, maximising
the fluxes of sources measured in the corresponding sky-subtracted images. We demonstrate for
data tuned to a central wavelength of 7615~$\rm\AA$ that galaxy fluxes in the new sky-subtracted 
image are $\sim37\%$ higher, versus a sky-subtracted image from existing methods for OSIRIS tunable 
filter data.

\end{abstract}

\section{Introduction}
A Fabry-P\'erot interferometer, or etalon, is comprised of two reflecting plates
working in a collimated beam. For a specific incidence angle of incoming light,
the etalon transmits light of wavelength $\lambda$ in a circular pattern of radius $r$ 
around the optical centre.  The range of wavelengths transmitted by the filter is
adjusted by changing the separation between the reflecting plates.

Tunable filter instruments (TFs), often built with Fabry-P\'erot interferometers, are 
proving to be a flexible and cost-effective implementation of spectrophotometry. The ability to 
precisely tune to an unlimited number of wavelengths in a specified interval circumvents 
the need to purchase arbitrary narrow-band filters (Bland-Hawthorn \& Jones 1998).
TFs are suitable for studies of emission and absorption lines in any redshift window, and 
they yield higher resolution ($R\sim500$) than low-resolution grisms (Gonz{\'a}lez et al. 2014). 
However, the varying wavelength across the field of view makes data from TF instruments
challenging to deal with.  Background sky emission can be highly variable across an image in which
bright OH sky emission lines appear as prominent rings (see Section~\ref{ssubtract} for an example). 
Full utilisation of TF data requires a precise wavelength calibration and robust means of subtracting 
the complicated sky pattern.

In this paper, we discuss refinements to the wavelength calibration and sky subtraction for TF data from 
the red mode on the Optical System for Imaging and low Resolution Integrated Spectroscopy instrument 
(OSIRIS, Cepa et al. 2013a,b) on the 10.4~m Gran Telescopio Canarias (GTC) telescope. We specifically 
consider data of emission line galaxies from the OSIRIS Mapping of Emission-line Galaxies in A901/2 
(OMEGA) survey (Chies-Santos et al. 2015) in the Space Telescope A901/2 Galaxy Evolution Survey (STAGES) 
field (Gray et al. 2009). Our discussion is generalisable to similar TF instruments. We summarise the 
most important properties of the data in Section~\ref{data}. Sections~\ref{swavecal} and \ref{ssubtract} 
address the wavelength calibration and sky subtraction, respectively.

\section[]{The OMEGA Survey}\label{data}
Here, we briefly summarise the survey design and relevant data acquisition details of OMEGA. For the
complete details, see Chies-Santos et al. (2015). OMEGA is based on a 90$\,$h ESO/GTC Large Programme 
allocation (PI: A. Arag\'on-Salamanca). It was designed to yield deep, spatially resolved emission-line 
images and low-resolution spectra covering the H$\alpha$ and {\rm [N\textsc{ii}]} lines for galaxies in the STAGES 
supercluster. 

An exposure time of 600$\,$s 
was adopted to achieve an $\rm{S/N} \geq 10$ in the line flux
for galaxies with $17 \leq R \leq 23.5$ (Vega) continuum magnitudes. Deblending the H$\alpha$ 
and {\rm [N\textsc{ii}]} lines required a TF full width at half maximum (FWHM) bandwidth of $14\,\rm{\AA}$ and a
wavelength sampling of $7\,\rm{\AA}$.  The 7615--7734$\,\rm\AA$ wavelength range was covered in 18
increments to probe the full cluster velocity range.  Twenty telescope pointings were
used to map 0.18$\,$deg$^2$ of the supercluster. 

Observations were taken over three observing seasons between 2012 and 2014. Clear sky conditions were
required, and the moonlight was grey or dark. Different exposures for a given field were not 
necessarily observed in the same night or under consistent environmental conditions (e.g., temperature,
humidity, seeing). The
median seeing was $\sim0\farcs9$, and always $\leq1\farcs2$. Additional details concerning these
observations and the data reduction are provided in Chies-Santos et al. (2015).

\section[]{Wavelength Self-Calibration}\label{swavecal}
Gonz{\'a}lez et al. (2014) show that the radial dependence of wavelength for the 
OSIRIS red TF is given by the expression

\begin{equation}
\lambda = \lambda_0 - 5.04 r^2 + a_3(\lambda)r^3
\end{equation}
where 
\begin{equation}
a_3(\lambda) = 6.0396 - 1.5698\times 10^{-3} \lambda + 1.0024\times 10^{-7} \lambda^2,
\end{equation}
$\lambda_0$ is the effective wavelength at the optical centre (i.e., the wavelength to which the TF
is tuned), $r$ is measured in arcmin, and wavelengths are measured in $\rm\AA$. After applying the 
above calibration to our data, we still found significant wavelength offsets between
the spectra of galaxies imaged independently in partially overlapping fields. The magnitude of 
the offsets varied from field to field, but it was in general enough to affect flux calibration and
velocity measurements.

Assuming the radial dependence of wavelength in Equation (1) is correct (which we will test later in this
section), we attempt to update the $\lambda_0$ term based on the positions of sky rings 
in the images. Adjusting $\lambda_0$ in this way essentially corrects for instrument tuning inaccuracies.

\begin{figure}
\begin{center}
\scalebox{0.43}{\includegraphics{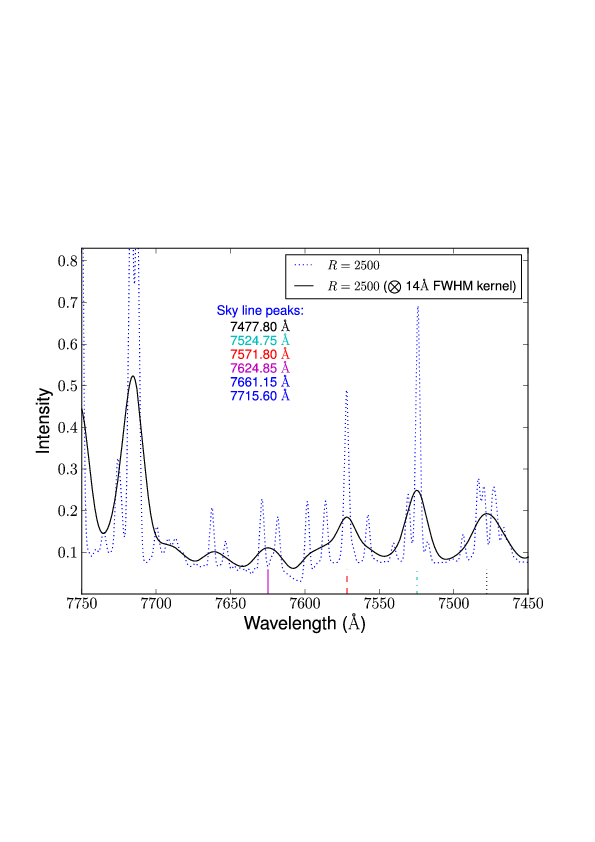}}
\caption{
The dashed curve is an intermediate-resolution ($R=2500$) sky spectrum from the GTC.  The solid curve results
after convolving the dashed curve with a Gaussian kernel having a FWHM equal to the resolution of our data 
(14~$\rm\AA$). The peaks of the sky lines in the convolved spectrum are measured as the local maximum around 
each peak. The colour coding and reversed wavelength scale make for easy comparison with Figures~\ref{skymap} and 
\ref{skyspectrum}.
\label{osirissky}}
\end{center}
\end{figure}

The high-resolution ($\rm R\approx35,000$ at 7000~$\rm\AA$) spectral atlas of Osterbrock et al. (1996)
shows multiple OH emission lines populate the spectral range of our observations.
We therefore simulate how the sky spectrum should look given our chosen TF bandwidth (14~$\rm\AA$).
We convolve an OSIRIS sky spectrum of resolution higher than our data with a 14~$\rm\AA$ FWHM Gaussian kernel. Figure~\ref{osirissky} shows the original sky spectrum
and the result of the convolution; central wavelengths of the sky lines in the low resolution spectrum
are measured simply as the local maxima of the peaks. 
The relative strengths of the night sky emission lines are known to vary with time, and this 
will affect the adopted convolved wavelength of the blended lines, limiting the accuracy of the 
wavelength calibrations. To evaluate the variability of this effect, we also convolved
an independent sky spectrum taken with the European Southern Observatory's Ultraviolet and Visual 
Echelle Spectrograph (UVES). All sky lines obtained after convolving the UVES spectrum agree to 
within $\pm0.14$~$\rm\AA$ of those from the convolved OSIRIS sky spectrum.

We generate sky spectra for every exposure to compare with the sky spectrum in Figure~\ref{osirissky} (see 
Chies-Santos et al. 2015 for the observing strategy). For the
sky background we simply use an intermediate-step frame from the OSIRIS Offline Pipeline Software (OOPS, 
Ederoclite 2012) that has been through all processing steps (overscan subtraction, bias subtraction, flat fielding)
except sky subtraction. (Alternatively, one could also directly use the sky models discussed in 
Section~\ref{ssubtract}).

The sky images are converted from Cartesian $x-y$ coordinates to polar $r-\theta$ coordinates, 
where $r$ is the distance of a pixel to the optical centre and $\theta$ is the angle from the image 
$y$-axis. The conversion of Cartesian to polar coordinates is made by backward mapping. A grid in 
polar coordinates with the desired resolution in $r-\theta$ is initialised, and then to each $r-\theta$ 
pixel the intensity at the corresponding Cartesian pixel is assigned.  We have used a resolution of 
1~pixel ($0\farcs25$)
in $r$ and 1 degree in $\theta$.  
Note, we adopt the optical centre reported by OSIRIS handbook\footnote{http://www.gtc.iac.es/instruments/osiris} 
of $X_0=772$, $Y_0=976$ (CCD1) and $X_0=-35$, $Y_0=976$ (CCD2).  

\begin{figure}
\centering
\begin{subfigure}
    \centering
    \scalebox{0.50}{\includegraphics[trim = 5mm 20mm 10mm 2mm, clip]{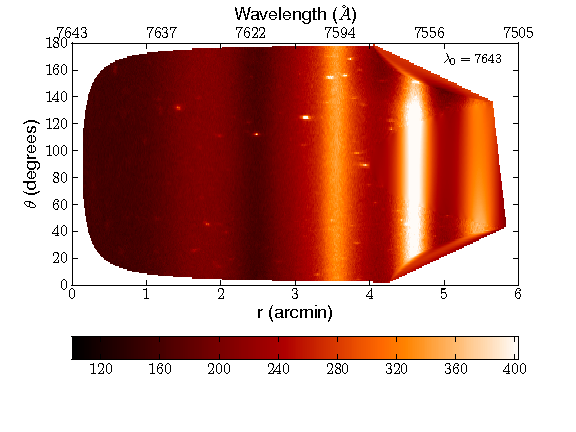}}
    \caption{This image is a two-dimensional sky map after conversion to polar coordinates. The
wavelength at the centre of the image is approximately 7643~$\rm\AA$, and the actual
wavelength probed declines from the centre outward. The vertical columns of relatively
higher intensity correspond to sky emission rings. These columns show no systematic
tilt and demonstrate the sky emission has circular symmetry. \label{skymap}}
\end{subfigure}
\begin{subfigure}
    \centering
    \scalebox{0.50}{\includegraphics[trim = 5mm 2mm 10mm 2mm, clip]{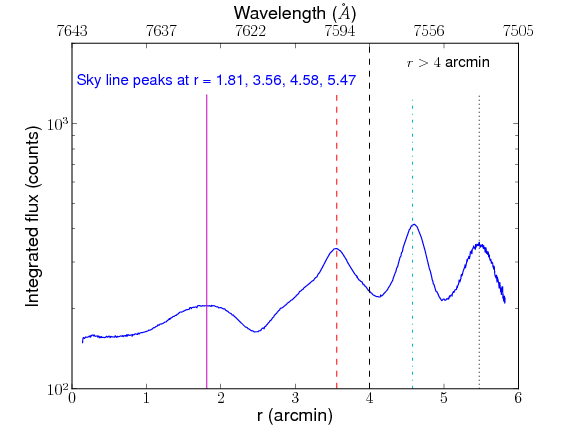}}
    \caption{This one-dimensional spectrum is the result of median collapsing across $\theta$ the sky map in
Figure~\ref{skymap}. The peaks correspond to sky lines at 7624.85, 7571.80, 7524.75,
and 7477.8~$\rm\AA$. The latter two lines are beyond a radius of 4 arcmin and are not used in the
recalibration.\label{skyspectrum}}
\end{subfigure}
\end{figure}

The example transformation in Figure~\ref{skymap} shows the sky emission rings become vertical 
columns in the $r-\theta$ plane. We have checked that there is no systematic change in the column centres 
(i.e., tilt) with $\theta$. This means the sky emission rings have circular symmetry and that we can accurately 
characterise their centre with a single measure after collapsing the $r-\theta$ plane in $\theta$.

The two-dimensional images are collapsed into one-dimensional spectra by taking the median across all 
$\theta$ at a given $r$. The example in Figure~\ref{skyspectrum}
shows a sky spectrum for an image where the central wavelength is approximately 7643~$\rm\AA$. Comparing
Figure~\ref{skyspectrum} to Figure~\ref{osirissky}, while noting the central wavelength of 7643~$\rm\AA$,
implies the visible sky lines in the spectrum correspond to wavelengths 7624.85, 7571.80, 7524.75, 
and 7477.8~$\rm\AA$. The central wavelength of each sky line in a one-dimensional spectrum is measured simply 
as the local maximum emission.

Reliable observations in the OSIRIS red TF data are limited to a circular field of view of 
radius 4 arcmin (960 pixels); beyond 4 arcmin, there can be contamination by other orders. 
For all sky lines at radii less than 4 arcmin from the optical centre, we measure the expected 
wavelengths of the sky lines using Equation (1). The average offsets between the actual and 
expected sky line wavelengths are the requisite adjustments to $\lambda_0$ for a single exposure.
As a simplification, we calculate average adjustments to $\lambda_0$ as a function of tuning wavelength
and field. The adjustments were typically $\sim5$~$\rm\AA$, but they were as high as $\sim$11~$\rm\AA$ 
in some cases.

\begin{figure*}
\centering
\subfigure[]{
    \centering
    \includegraphics[height=3in]{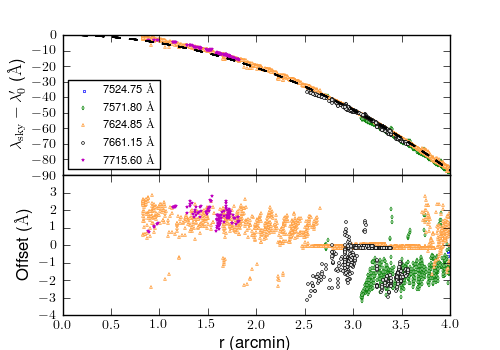}
    \label{testcal}}
\hspace{1cm}
\subfigure[]{
    \centering
    \includegraphics[height=3in]{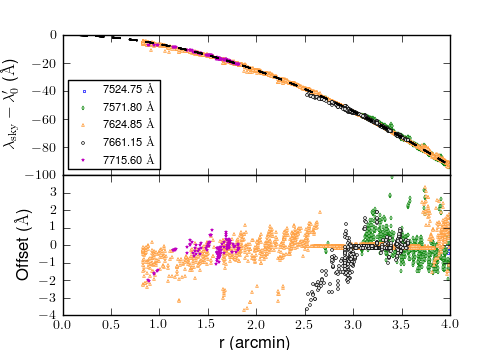}
    \label{testcalnew}}
\caption{In panel~\subref{testcal} (top), the difference in the sky line wavelength ($\lambda_{\rm sky}$) and the 
recalibrated tuning wavelength ($\lambda_0^\prime$) is plotted against radius from the optical centre. The dashed 
line is the wavelength calibration derived by Gonz{\'a}lez et al. (2014). The bottom panel shows the corresponding 
residuals. The bands of near zero residual are from exposures with only one sky line at $r<4$ arcmin. 
Panel~\subref{testcalnew} is similar but instead uses the result of the iterative refitting procedure discussed in 
Section~\ref{swavecal}. }
\end{figure*}

In Figure~\ref{testcal} we test whether the updated calibration still follows the relation
from Gonz{\'a}lez et al. (2014, our Equation 1). We plot $\lambda_{\rm sky} - \lambda_0^\prime$, 
where $\lambda_{\rm sky}$ is the sky line wavelength and $\lambda_0^\prime$ is the recalibrated 
tuning wavelength, for sky lines within 4~arcmin of the optical centre. The line is the 
relation from Gonz{\'a}lez et al. (2014).  Offsets of $\sim2-3$~$\rm\AA$ from the calibration
are common. Gonz{\'a}lez et al. (2014) assumed the CS-100 Fabry-P\'erot controller in 
OSIRIS is strictly linear in its gap spacing-control variable (Z) relation, and attributed all 
non-linearities to phase dispersion effects in the dielectric coatings of the etalons. The 
discrepancy we measure here may be an indication that the assumption of linearity is not 
strictly true.

While the relation from Gonz{\'a}lez et al. (2014) is accurate
over a large wavelength range, systematic offsets up to 3~$\rm\AA$ can occur in certain narrow
wavelength ranges. One can proceed with this level of disagreement if it does not affect the
accuracy of the science, e.g., H$\alpha$-based star-formation rates.  For redshift determinations,
however, a 2~$\rm\AA$ error (93 km/s) is large.

We propose an iterative method to refit the wavelength calibration and reduce the typical
error down to 1~$\rm\AA$. In our data set, most (89\%) exposures with more than one sky line 
inside 4~arcmin of the optical centre contain the sky line at 7624.85~$\rm\AA$. We focus on 
these exposures and perform an iterative procedure.

\begin{itemize}
\item Calculate offsets relative to the Gonz{\'a}lez et al. (2014) calibration for sky line
7624.85~$\rm\AA$ as a function of tuning wavelength setting ($\lambda_0$) and field.
\item Apply these offsets to the remaining sky lines at wavelengths other than 7624.85~$\rm\AA$.  
To these sky lines, fit the same functional form justified by Gonz{\'a}lez et al. (2014), but 
with more free parameters for better agreement, namely

\begin{equation}
\lambda = \lambda_0 + a_2 r^2 + a_3(\lambda)r^3
\end{equation}
where
\begin{equation}
a_3(\lambda) = a_{3,0} - 1.5698\times 10^{-3} \lambda + 1.0024\times 10^{-7} \lambda^2.
\end{equation}
The free parameters are $a_2$ and $a_{3,0}$, which are fixed to -5.04 and 6.0396 by Gonz{\'a}lez et al. (2014).
\item Iterate until the model converges with the sky lines at 7624.85~$\rm\AA$. In subsequent 
iterations, the offsets for the 7624.85~$\rm\AA$ sky line are calculated relative to the newly fit 
model in the above step. 
\end{itemize}

Figure~\ref{testcalnew} shows the result after 10 iterations of this procedure.  The fit is noticeably 
better overall (although residuals for a few measurements slightly worsen). Most (85.5\%) of the individual 
sky line measurements agree to within 1~$\rm\AA$ of the new calibration. This is an improvement over 
Figure~\ref{testcal} where that percentage is only 50.7\% for the Gonz{\'a}lez et al. (2014) solution.

Using the new calibration from Figure~\ref{testcalnew}, we calculated average corrections to $\lambda_0$ 
for each combination of wavelength setting and field in our data.  
Correcting the central wavelengths with the offsets calculated from the sky lines in this way yields a 
wavelength calibration accurate to within $0.5$~$\rm\AA$ for most (76\%) of the individual image frames, 
and accurate to within $1$~$\rm\AA$ for 95\% of the frames.  The 1~$\rm\AA$ level of accuracy is a success 
considering it is $\sim7\%$ of the 14~$\rm\AA$ instrumental resolution.
The calculated corrections do not vary in any systematic way with date of observation,
ambient temperature, or humidity.  The mean offset does vary weakly with $\lambda_0$.  The
average offset declines from $2$~$\rm\AA$ at $\lambda_0=7620$~$\rm\AA$ to $-2.7$~$\rm\AA$ at 
$\lambda_0=7734$~$\rm\AA$.

It is important to point out that OSIRIS uses a non-standard phase correction scheme that
may affect the generalisability of this method to very different Fabry-P\'erot interferometers.
Additional prerequisites for the application of this method are that the data be circularly 
symmetric around the optical centre and that wavelength dependence on detector position 
be radially symmetric. Our wavelength recalibration benefited from having a common sky 
line across most exposures; not having this may produce poorer results. 
This technique's 
accuracy may further be limited by variation in sky line relative intensities, which can perturb 
the effective peak positions of the blended sky lines in the low-resolution OSIRIS sky spectra.
For our TF bandpass ($\sim14$~$\rm\AA$) and spectral range, the variability is small 
($\pm0.14$~$\rm\AA$), but it may be worse in other instances.

\section[]{Sky Subtraction}\label{ssubtract}
\begin{figure*}
\centering
\subfigure[]{
    \centering
    \includegraphics[height=4.0in]{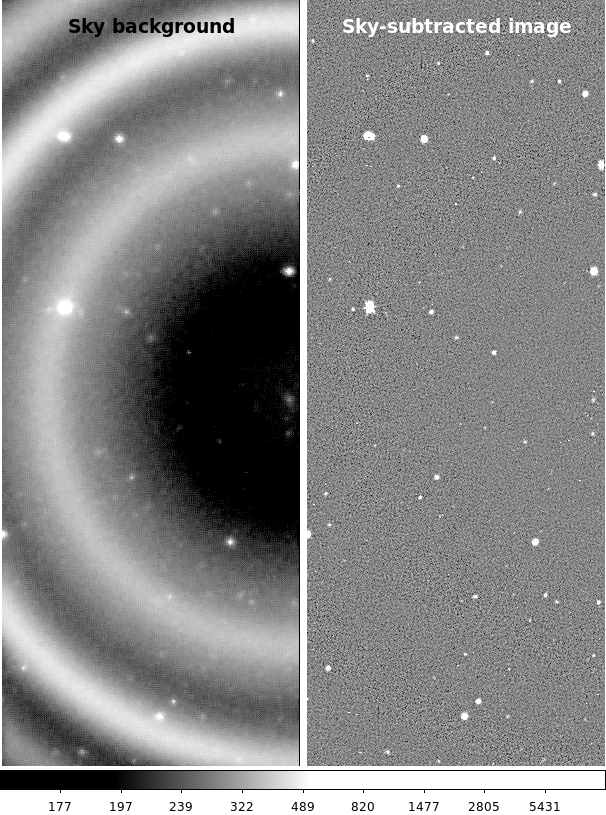}
    \label{oopsExample}}
\hspace{4cm}
\subfigure[]{
    \centering
    \includegraphics[height=4.0in]{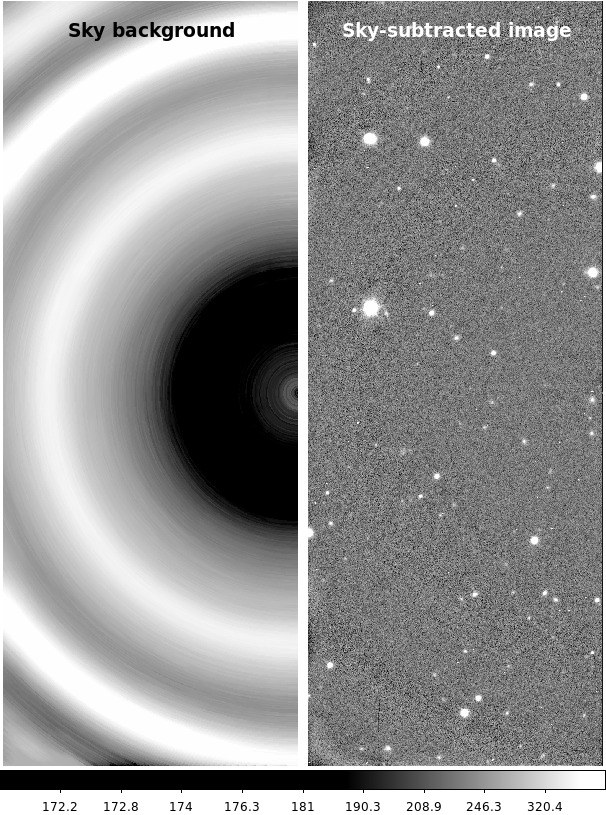}
    \label{filtExample}}
\caption{For a representative exposure, panel~\subref{oopsExample} shows the sky background measured by OOPS (left), 
and the corresponding sky-subtracted image (right). The wavelength at the optical centre is approximately 
7615~$\rm\AA$. For the same exposure, panel~\subref{filtExample} shows an example of the sky map generated by the 
filtering technique outlined in Section~\ref{ssubtract} (left) and the corresponding sky-subtracted image (right).  
Median filtering was performed with $dr$ set to 3 pixels ($0\farcs75$), a total filter size $dr d\theta$ of 170 
pixel-radians, and iterative $3\sigma$ clipping. The colour bars show the colour stretch and are similar in both 
panels.}
\end{figure*}

In this section we overview a new sky subtraction technique for TF data.
The current sky subtraction in OOPS works by artificially dithering
images on a 3-by-3 pixel grid (Ederoclite 2012).  Median combining the dithered 
images produces an estimate of the sky background. Visual inspection of the sky
maps, such as the one shown in Figure~\ref{oopsExample} tuned to a central wavelength
of 7615~$\rm\AA$, shows that the outer envelopes of galaxies remain in the sky background.
Applying this sky background oversubtracts and eliminates the real outer envelopes
of galaxies.

We have developed an alternate approach to sky subtraction that mitigates the problem
of source oversubtraction. The basic idea is to remap the flat-fielded images from 
Cartesian to polar coordinates (Section~\ref{swavecal}) and then run a median filter, 
in polar coordinates, along circular arcs of grouped radii, to filter out sources and 
leave behind an estimate of the sky background. 

Circular symmetry in the background sky emission means that the sky changes quickly
in $r$ but is relatively stable across $\theta$ at a given $r$ (Figure~\ref{skymap}). However, 
sky subtraction is not as straightforward as converting a collapsed one-dimensional sky spectrum 
(e.g., Figure~\ref{skyspectrum}) into a two-dimensional circularly-symmetric sky model. Because in some cases the sky varies 
as a slow function of $\theta$ ($5$--$10$\% amplitude) due to imperfect flatfielding, this simplified approach
is less effective at removing broadened sky line emission than the method we will adopt.

The simplest application of a median filter is to filter over $\theta$ at a given $r$ with a fixed 
window size (as measured in polar image pixels). This is also not recommended. Sources at small 
distances from the optical centre are not effectively removed because, for a fixed window size in 
polar coordinates, the number of pixels from the original image that contribute to the median becomes very small 
at small radii. A significant improvement in filtering performance 
is achieved by fixing the median filter window to contain an approximately constant number of pixels from the 
original image. Thus, for a given filtering window $dr d\theta$, the number of polar image pixels in the filter varies 
inversely with radius, and this ensures a sufficient number of pixels from the original image are filtered over at 
small radii.

As we show below, changing the width of $d\theta$ has little effect on the outcome 
of the sky subtraction.  Changing $dr$, however, has a more significant effect. Increasing 
$dr$ incorporates more image pixels into the filter, improving robustness and increasing the
speed of the filtering. Raising $dr$ too high yields a sky map with too low spectral resolution that 
introduces artifacts in the sky-subtracted images. As a compromise, we have chosen to set 
$dr$ to 3 pixels ($0\farcs75$, 0.57~$\rm\AA$ at $r=4$ arcmin), small enough to be beneficial
but not problematic. 

Because the dependence of $\lambda$ on $r$ is approximately quadratic, 
a more sophisticated approach would be to change $dr$ according to quadratically varying bins 
such that each radial bin represents a constant $\Delta\lambda$ decrement in wavelength. We have 
tested this binning pattern with our data. Even if we make the radial bins 1~pixel wide at large radii,
the radial bins near the optical centre become so wide that often include many sources which are not effectively removed by
the median. This leaves behind residual background light in the 
sky-subtracted images that would bias the photometry of centrally located sources. Other, more complex, radial 
binning patterns could also be applied, but we did not find they yield any significant improvement over our method. 
Nevertheless, while this increased complexity is not necessary with our data, it may prove useful with different 
data sets.

As a further refinement, we apply sigma clipping to prevent the extended halos of bright 
sources from being smeared into the background sky, which causes local oversubtraction of the sky
and leads to dark halos around galaxies. We have found that iteratively clipping to 
$3\sigma$ for 5 iterations is sufficient to nullify this effect.

\begin{figure*}
\scalebox{0.40}{\includegraphics{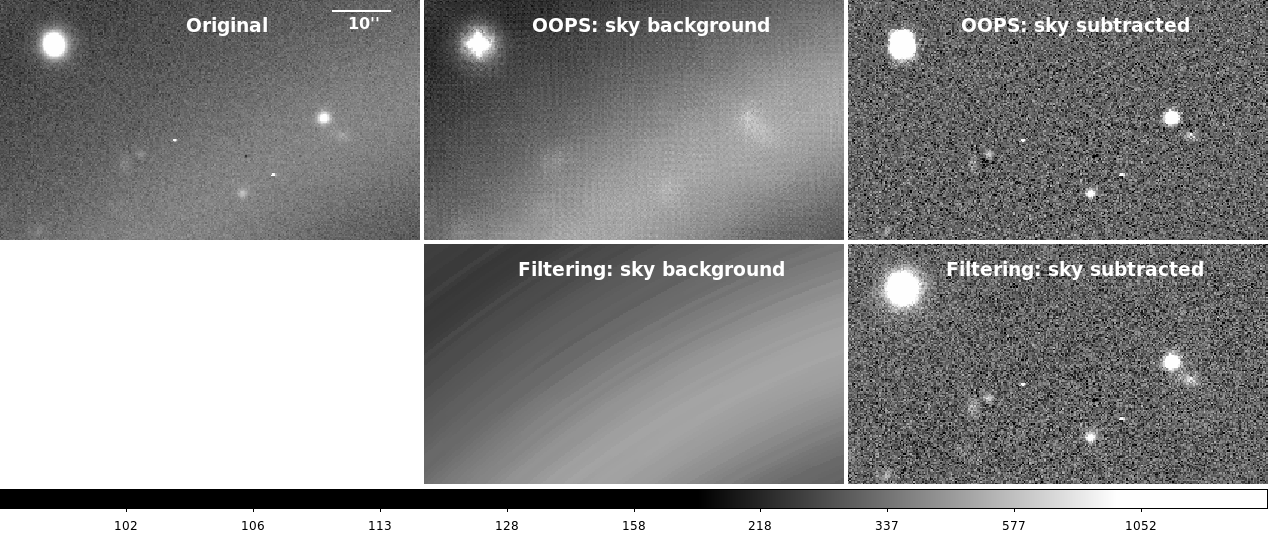}}
\caption{
A magnified view of Figures~\ref{oopsExample} and \ref{filtExample} showing a subregion containing both bright 
extended sources and compact faint sources. The top row shows the original flat-fielded image still containing 
the sky, the sky background from OOPS, and the final OOPS sky subtraction. The second row shows the sky 
background inferred from filtering and the resulting sky-subtracted image. The colour bar shows the colour 
stretch and applies to all panels.
\label{zoomin}}
\end{figure*}

Figure~\ref{filtExample} shows the results of applying the filtering to the same 
exposure highlighted in Figure~\ref{oopsExample}. For this example, we have fixed 
$dr$ to 3 pixels and the total filter size $dr d\theta$ to 170 pixel-radians, 
corresponding to a filter size of 20 degrees at $r=2$ arcmin. The 
sky background is much smoother than what is provided by the current algorithm in OOPS
(Figure~\ref{oopsExample}, left-hand panel), and the sky-subtracted sources in the
right-hand panel of Figure~\ref{filtExample} are clearly brighter than in
Figure~\ref{oopsExample}. Binning in $r$, however, yields sky models with band-like 
artifacts of width $dr$ (see Figure~\ref{zoomin}, row 2, column 2).  The change in 
brightness between bands is typically $\sim$1-2\% of the sky background, smaller than 
the background shot noise (3--4\%).

Figure~\ref{zoomin} gives a magnified view of the sky subtraction
in an image subregion containing bright extended objects as well as small faint sources. 
The sky subtracted-images in the third column emphasise that galaxy light sacrificed
by the standard OOPS algorithm is retained with median filtering.

To quantify how much light was gained with the revised sky subtraction procedure, we 
measured the fluxes of sources in the sky-subtracted images using the methodology of 
Chies-Santos et al. (2015).  
Fluxes of galaxies in the right-hand panel of Figure~\ref{filtExample}
are higher by a median of $\sim37\%$ versus the OOPS image in Figure~\ref{oopsExample}. 
This boost in flux is not strongly sensitive to either $dr$ or the total filter size 
$dr d\theta$. We repeated this test for filters with $dr$ values of 1--5 pixels (at 
fixed $dr d\theta$ of 170 pixel-radians), as well as for $dr d\theta$ values of 10--30 
degrees at $r=2$ arcmin (with $dr=3$ pixels). The resulting median flux ratios deviate 
by $\sim1-3\%$ from the $\sim37\%$ value obtained with the adopted fiducial filter 
parameters ($dr = 3$ pixels, $dr d\theta = 170$ pixel-radians).

This filtering technique is implemented in Python using standard libraries, including
\texttt{Astropy} (Astropy Collaboration et al. 2013).  We are assisting with the 
incorporation of this technique into
OOPS so that all users of OSIRIS will have access (Ederoclite, private communication).

\section[]{Summary}\label{ssummary}
In this letter, we have used data from the red TF mode on OSIRIS to demonstrate
new techniques for wavelength calibration and sky subtraction. Central to our
methodology is the use of polar coordinates, which simplifies matters when the TF 
data is circularly symmetric. In Section~\ref{swavecal}, 
we outlined a technique for wavelength recalibration using OH sky emission rings. This 
approach increases the accuracy of the absolute wavelength calibration from $\sim5$~$\rm\AA$
to 1~$\rm\AA$, or $\sim7\%$ of the instrumental resolution. 
In Section~\ref{ssubtract}, we presented a new method to 
estimate the sky background by median filtering in polar coordinates.  The merit of
this approach is that of light from the extended halos of emission-line galaxies does
not contaminate the background sky maps. Sources in the associated sky-subtracted images
will likely be significantly brighter than how they would appear in images based on the 
current sky-subtraction algorithm in the OSIRIS reduction software pipeline, OOPS. Future 
OSIRIS/OOPS users will benefit from this sky subtraction method.

\section*{Acknowledgments}
Based on observations made with the Gran Telescopio Canarias, installed in the Observatorio del Roque de los Muchachos of the Instituto de Astrof\'isica de Canarias, in the island of La Palma. The GTC reference for this programme is GTC2002-12ESO. Access to GTC was obtained through ESO Large Programme 188.A-2002.

\label{lastpage}


\begin{thebibliography}{}
\bibitem[Astropy Collaboration et al.(2013)]{2013A&A...558A..33A} Astropy Collaboration, Robitaille T.~P., Tollerud E.~J., et al., 2013, \aap, 558, A33 

\bibitem[]{} Bland-Hawthorn J., 2000, in van Breugel W., Bland-Hawthorn J., eds, ASP Conf. Ser. Vol. 195, Imaging the Universe in Three Dimensions. Astron. Soc. Pac., SanFrancisco, p. 34

\bibitem[]{} Chies-Santos A. et al., 2015, MNRAS, in press

\bibitem[]{} Cepa J., Bongiovanni A., P\'erez Garc\'ia A. M., et al. 2013, Highlights of Spanish Astrophysics VII, 868

\bibitem[]{} Cepa J. 2013b, Revista Mexicana de Astronomia y Astrofisica Conference Series, 42, 77

\bibitem[]{} Ederoclite A., \textit{The OSIRIS Offline Pipeline Software (OOPS)}, 2012

\bibitem[Gonz{\'a}lez et al.(2014)]{2014MNRAS.443.3289G} Gonz{\'a}lez J.~J., Cepa J., Gonz{\'a}lez-Serrano J.~I., \& S{\'a}nchez-Portal M.\ 2014, \mnras, 443, 3289 

\bibitem[Gray et al.(2009)]{2009MNRAS.393.1275G} Gray, M.~E., Wolf, C., Barden, M., et al.\ 2009, \mnras, 393, 1275 

\bibitem[Osterbrock et al.(1996)]{1996PASP..108..277O} Osterbrock, D.~E., Fulbright, J.~P., Martel, A.~R., et al.\ 1996, \pasp, 108, 277 

\end{thebibliography}
\end{document}